\documentclass[prl,twocolumn,superscriptaddress,showpacs,amsmath,amssymb]{revtex4-1}
\usepackage{graphicx}% Include figure files
\usepackage{dcolumn}% Align table columns on decimal point
\usepackage{bm}% bold math
\usepackage{color}
\usepackage{epsfig,float,afterpage,amssymb,wrapfig,psfrag}
\usepackage{subfigure}

\begin{document}

\title{Quantum phases of the Shastry-Sutherland Kondo lattice:\\
implications for the global phase diagram of heavy fermion metals}
\author{J.~H.~Pixley}
\affiliation{Department of Physics \& Astronomy, Rice University,
Houston, Texas, 77005, USA}
\author{Rong~Yu}
\affiliation{Department of Physics and Beijing Key Laboratory of Opto-electronic Functional Materials and Micro-nano Devices, Renmin University of China, Beijing 100872, China}
\affiliation{Department of Physics \& Astronomy, Rice University,
Houston, Texas, 77005, USA}
\author{Qimiao Si}
\affiliation{Department of Physics \& Astronomy, Rice University,
Houston, Texas, 77005, USA}

\date{\today}

\begin{abstract}
Considerable recent theoretical and experimental efforts have been devoted to the study of quantum
criticality and novel phases of antiferromagnetic heavy-fermion metals. In particular, quantum
phase transitions have been discovered in 
heavy-fermion compounds with geometrical frustration.
These developments have motivated us to
study the competition between the RKKY and Kondo interactions on the Shastry-Sutherland lattice.
We determine the zero-temperature phase diagram as a function of magnetic frustration and Kondo coupling
within a slave-fermion approach.
Pertinent phases include the 
valence bond solid and heavy Fermi liquid.
In the presence of antiferromagnetic order,
our zero-temperature phase diagram is remarkably similar to
the global phase diagram proposed earlier based on general grounds.
We discuss the implications of our results for the experiments on
Yb$_2$Pt$_2$Pb and
related
 compounds.
\end{abstract}

% PACS:
% 71.10.Hf Non-Fermi-liquid ground states, electron phase diagrams and
% phase transitions in model systems
% 71.27.+a Strongly correlated electron systems; heavy fermions
% 75.20.Hr Local moment in compounds and alloys; Kondo effect, valence
% fluctuations, heavy fermions (see also 72.15.Qm Scattering
% mechanisms and Kondo effect)
% 05.10.Cc Renormalization group methods

\maketitle

Geometrical frustration in insulating quantum antiferromagnets can lead to a variety of quantum phases,
such as valence bond solids (VBS) and quantum spin liquids~\cite{Balents}.
Recent studies have discovered intriguing properties
in a growing list of metallic systems with local magnetic moments residing on frustrated lattices.
In these heavy fermion compounds,
the interplay of Kondo screening and magnetic frustration may give rise to entirely new ground states and
quantum phase transitions~\cite{Si}.
 For example, the compounds Yb$_2$Pt$_2$Pb~\cite{Kim} and CePd$_{1-x}$Ni$_x$Al~\cite{Fritsch}
 have spin-$1/2$ local moments located
 on the Shastry-Sutherland and Kagome lattices respectively.
 Likewise, both YbAgGe~\cite{Morosan}
  and YbAl$_3$C$_3$~\cite{Khalyavin}
 feature triangular lattices.
All these compounds show an enhanced specific heat coefficient, implying a large effective mass
and the presence of the Kondo 
effect.

\begin{figure}[t]
\includegraphics[height=2.0in]{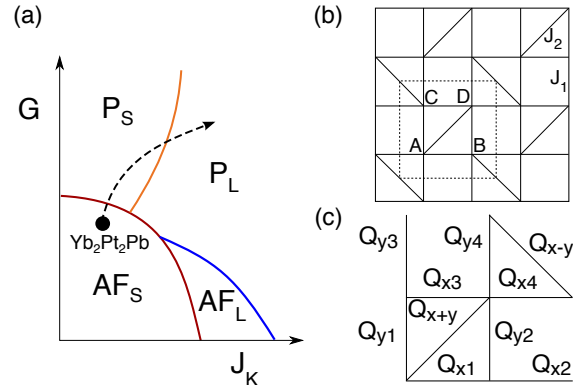}
\caption{ (color online).  (a) Proposed global phase diagram of heavy fermion metals.
Here, AF$_S$ and AF$_L$ refer to antiferromagnetic states without or with
static Kondo screening.
P$_L$ is the paramagnetic heavy Fermi liqud, and P$_S$ refers to
a paramagnetic phase without static Kondo screening.
Adapted from Ref.~\cite{Si2}.
We have sketched the proposed trajectory of Yb$_2$Pt$_2$Pb under magnetic field tuning.  
(b)  Shastry-Sutherland lattice,
denoting the Heisenberg exchange couplings $J_1$
on all the horizontal and vertical bonds, and $J_2$ along the diagonals.
The unit cell is the dashed square,
containing four sites $A,B,C,D$.  (c)
The bond singlet parameters.
}
\label{fig:model}
\end{figure}

General theoretical considerations of the competition between Kondo and RKKY interactions
have led to a proposal for the global phase diagram of heavy fermion metals
as a function of frustration or quantum fluctuations ($G$), and the Kondo coupling ($J_K$)~\cite{Si2,Coleman}; see Fig.~\ref{fig:model}(a).
This phase diagram incorporates not only antiferromagnetic (AF)
order, but also the physics of Kondo destruction \cite{LCP,JPCM,Senthil}.
From the Kondo-destroyed antiferromagnetic  phase (AF$_S$), the transition to the heavy fermi liquid phase
(P$_L$) could
take place directly (type I), via the spin-density-wave phase (AF$_L$) (type II), or through the Kondo-destroyed
paramagnetic phase ($P_S$) (type III). The heavy fermion compounds CeCu$_{6-x}$Cu$_x$,
YbRh$_2$Si$_2$ and CeRhIn$_5$ have shown strong evidence
for realizing
the type I transition \cite{Schroder,Friedeman,Shishido,Park06}. CePd$_3$Si$_{20}$,
which is cubic and therefore would have a smaller $G$, has properties consistent with a type II transition
\cite{Custers12}. Geometrical frustration is expected to enhance the quantum fluctuation parameter $G$,
raising the prospect of realizing a type III transition.
There is a recent surge of heavy-fermion materials that appear to be suitable for exploring
 this large-$G$ portion
of the global
phase diagram.
In particular, Yb$_2$Pt$_2$Pb and its homologues such as Ce$_2$Pt$_2$Pb~\cite{Kim},
featuring the geometrically-frustrated Shastry-Sutherland lattice,
may involve an intermediate VBS P$_S$ phase.

%%%%%%%%%%%%%%% Model fig %%%%%%%%%%%%%
%%%%%%%%%%%%%%% Model fig %%%%%%%%%%%%%

In this work, we study the effect of frustration on the
Kondo-Heisenberg model by considering it
on the Shastry-Sutherland lattice (SSL) \cite{Shastry}, as illustrated in Fig.~\ref{fig:model}(b).
%We determine the phase diagram as a function of frustration and Kondo
%coupling within a  slave-fermion approach. We
%identify and
%clarify the role of
%intermediate low energy states that break the
%symmetry of the SSL and exhibit partial Kondo screening (PKS).
%In addition, we show that the phase diagram we determine
%provides the first concrete realization of the
%global phase diagram [see Figs.~\ref{fig:model}(a) and \ref{fig:M-pd}] previously advanced based
%on general theoretical considerations.
The
%Kondo-Heisenberg
Hamiltonian
 %on the SSL
  is defined as
\begin{equation}
H = \sum_{(i,j),\sigma}t_{ij} ( c_{i\sigma}^{\dag}c_{j\sigma}  + \mathrm{h.c.})
+ J_K \sum_i {\bf S}_i\cdot{\bf s}_i^c + \sum_{(i,j)}J_{ij}{\bf S}_i\cdot{\bf S}_j
\label{eqn:ham}
\end{equation}
where $(i,j)$ denote the nearest neighbors (NN) and next nearest neighbors (NNN)
on the SSL as shown in Fig.~\ref{fig:model}(b).
The NN and NNN tight binding parameters for the conduction electrons,
denoted by $c_{i\sigma}$, are $t_1$  and $t_2$, respectively.
The spins of the conduction electrons are
${\bf s}_i^c = c_{i\alpha}^{\dag}(\bm{\sigma}_{\alpha\beta}/2)c_{i\beta}$ at site $i$,
where $\bm{\sigma}_{\alpha\beta}$ are the Pauli spin matrices.
They are coupled to spin-$1/2$ local moments, ${\bf S}_i$,
through an antiferromagnetic Kondo coupling $J_K$.
We have explicitly included the RKKY interactions,
incorporating $J_1$ and $J_2$, the NN and NNN terms respectively
[Fig.~\ref{fig:model}(b)].
The degree of frustration is measured
by the ratio $G=J_2/J_1$.  We represent the local moments using fermionic spinons~\cite{Auerbach},
$f_{i\sigma}$ such that ${\bf S}_i = f_{i\alpha}^{\dag} (\bm{\sigma}_{\alpha\beta}/2) f_{i\beta}$
with  a constraint $\sum_{\sigma}f_{i\sigma}^{\dag}f_{i\sigma}=1$ at each lattice site.
The spin-$1/2$ Heisenberg model on the SSL was extensively studied
(e.g., Refs.~\cite{Shastry,Chung,Lauchli,Miyahara}).
For $J_2/J_1>2$, it possesses an exact VBS ground state, where singlets form
across each disconnected diagonal bond~\cite{Shastry}.
Whereas for small $J_2/J_1$, the model has an AF  ground state~\cite{Chung,Lauchli}.
The transition between these two states has not been completely
determined ~\cite{Chung,Lauchli}.
The model in the presence of Kondo coupling was studied in some detail by
Bernhard {\it et al.} \cite{Bernhard}, and was also discussed qualitatively \cite{Coleman}.
As we will discuss, our work here reports the first complete analysis of the relevant
phases, and this is essential both in realizing the global phase diagram and in shedding
light on the experimentally observed PKS phase.

\emph{Large-$N$ limit}:
Generalizing the spin symmetry from $SU(2)$ to $SU(N)$,
we arrive at $H = \sum_{(i,j),\sigma}t_{ij} (c_{i\sigma}^{\dag}c_{j\sigma} + \mathrm{h.c.}) -
J_K/N \sum_i :B_i^{\dag}B_i:  -\sum_{(i,j)}(J_{ij}/N):D_{ij}^{\dag}D_{ij}:$
where $B_i = \sum_{\sigma}c_{i\sigma}^{\dag}f_{i\sigma}$, $D_{ij}
= \sum_{\sigma}f_{i\sigma}^{\dag}f_{j\sigma}$.
The sum now runs over $\sigma = 1,\dots,N$,
and the constraint becomes
$\sum_{\sigma}f_{i\sigma}^{\dag}f_{i\sigma}=N/2$. We have also used
$:\dots:$ to denote normal ordering.
The large-$N$ mean field Hamiltonian can be expressed as:
\begin{eqnarray}
H_{MF} &=& E-\sum_{(i,j),\sigma}( Q_{ij}^*f^{\dag}_{i\sigma}f_{j\sigma} + \mathrm{h.c.} )
+\sum_{i,\sigma}\lambda_if^{\dag}_{i\sigma}f_{i\sigma}
\nonumber
\\
&+&\sum_{(i,j),\sigma}t_{ij} (c^{\dag}_{i\sigma}c_{j\sigma} + \mathrm{h.c.})
-\sum_{i,\sigma}(b_i^*c^{\dag}_{i\sigma}
f_{i\sigma} + \mathrm{h.c.}).
\nonumber
\\
\label{eqn:Hmf}
\end{eqnarray}
We have used a Hubbard-Stratonovich transformation
decoupling $B_i$ and $D_{ij}$ in the Kondo singlet
and resonating valence bond (RVB) channels respectively
~\cite{Senthil,Coleman2}, and the constraint is enforced by $\lambda_i$.
The constant term is
$E/N=\sum_i\left(|b_i|^2/J_K-\lambda_i/2\right)+
\sum_{(i,j)}|Q_{ij}|^2/J_{ij}$.  The Kondo
parameter
$N b_i=J_K\langle B_i \rangle$ can be taken to be real by absorbing
its phase into the constraint field $\lambda_i$~\cite{Read}, whereas
the RVB parameters $N Q_{ij}=J_{ij}\langle D_{ij} \rangle$ are in general complex.

We solve Eq.~(\ref{eqn:Hmf})
by using a four-site unit cell,
where each site is
labeled by
$i\rightarrow({\bf r},X)$,
with $X=A,B,C,D$ marking the sublattice [see Fig.~\ref{fig:model}(b)], and
${\bf r}$ specifying a unit cell. We introduce Fourier transforms
per sublattice~\cite{Dodds} as $c_{{\bf r} X \sigma} =
1/\sqrt{N_u}\sum_{{\bf k}}e^{-i{\bf k}\cdot({\bf r}+\bm{\delta}_X)}c_{{\bf k} X \sigma}$,
where $\bm{\delta}_X$ points to each sub lattice $X$ from sub-lattice $A$.
Keeping the full generality of the four-site unit cell we introduce
sublattice dependent Kondo parameters and constraint fields
$b_X,\lambda_X$,
and use ten complex RVB
parameters $Q_{ij}$ as shown in Fig.~\ref{fig:model}(c).
These parameters are determined by solving the saddle-point equations self-consistently 
(see supplementary material~\cite{supp}).
We consider the metallic case $0<n_c<1$,
where $n_c=\frac{1}{4N_u}\sum_{i,\sigma} \langle c^\dagger_{i\sigma} c_{i\sigma}\rangle$ is the
filling of the conduction band.

\begin{figure}[t!]
\includegraphics[height=2.in]{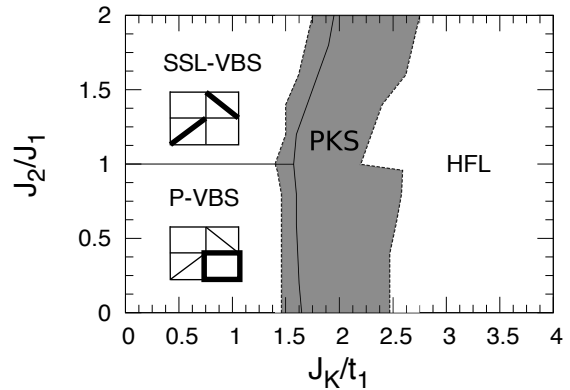}
\caption{(color online).  Large-$N$ phase diagram as a function of frustration ($J_2/J_1$)
and Kondo coupling ($J_K/t_1$), for a metallic filling $n_c=0.5$.
The phases are described in the main text.
The solid lines represent first-order transitions, and the dashed lines surrounding the grey area
locate the boundaries of the intermediate phases
that exhibit partial Kondo screening (PKS).}
\label{fig:phase-diagram}
\end{figure}

\begin{figure}[t!]
\begin{minipage}[b]{20pc}
\includegraphics[height=2.4in, angle=-90]{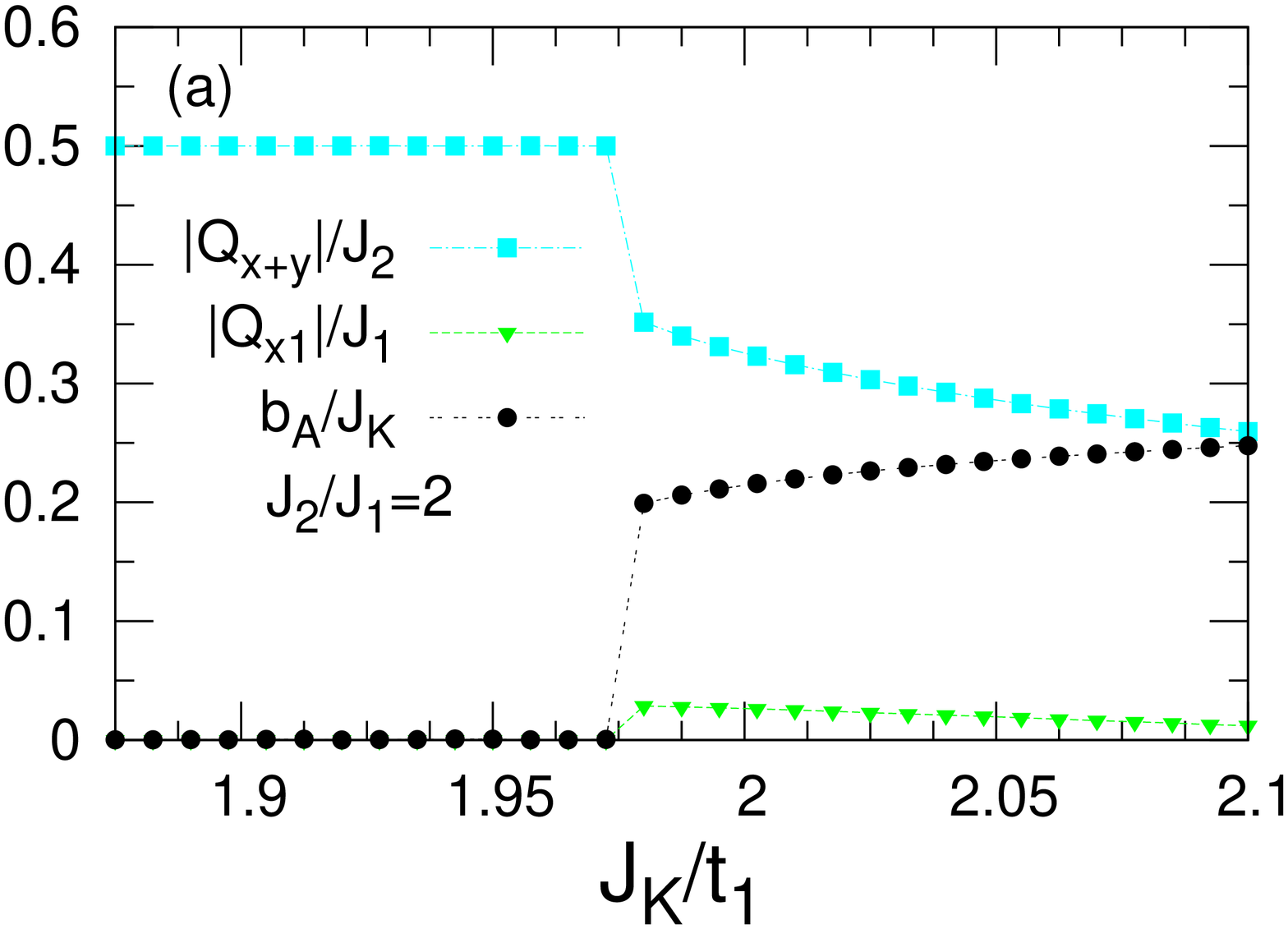}
\end{minipage}%\hspace{0.5pc}
\newline
\begin{minipage}[b]{10pc}
\includegraphics[height=1.7in, angle=-90]{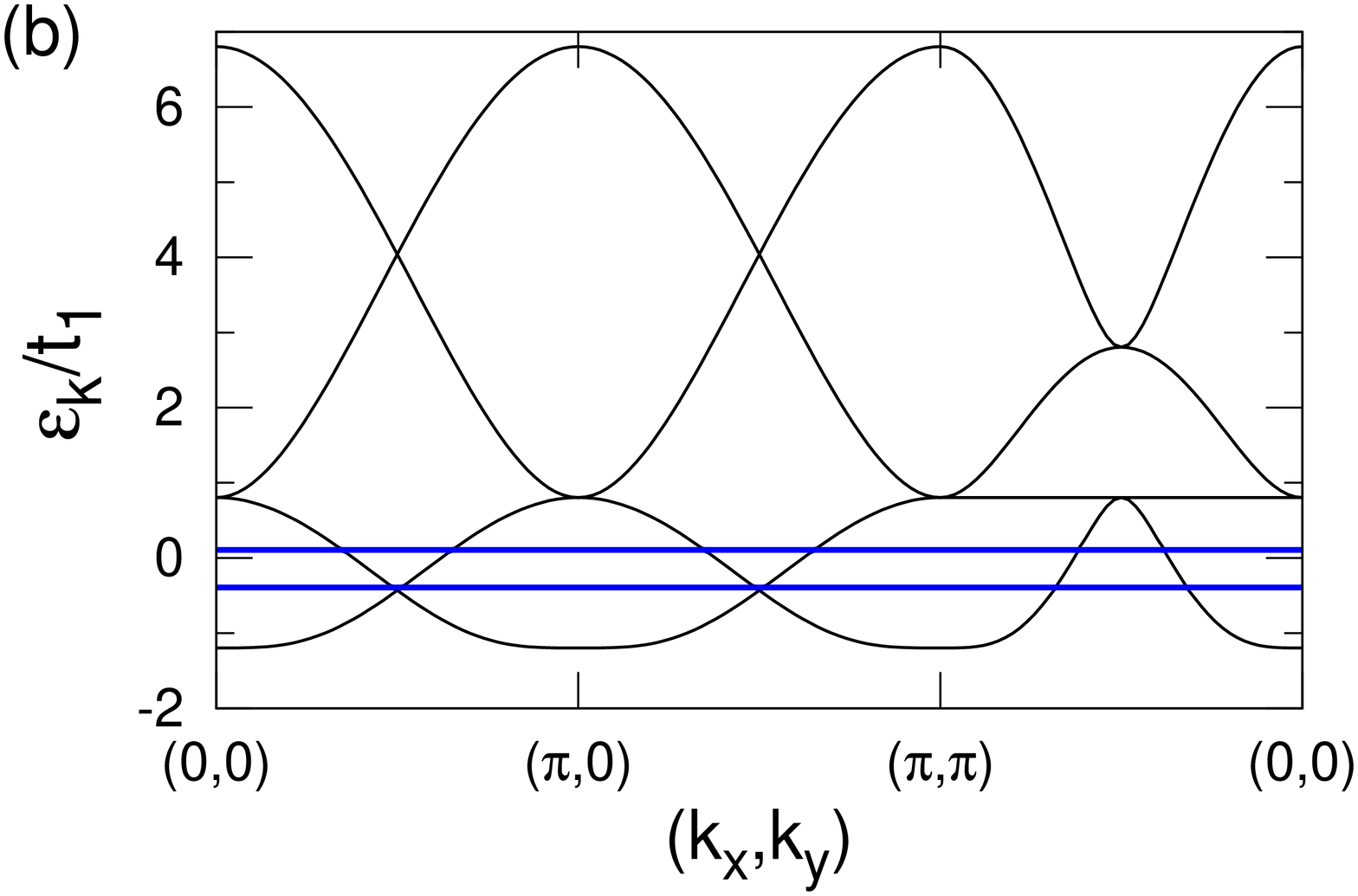}
\end{minipage}%\hspace{0.5pc}
\begin{minipage}[b]{10pc}
\includegraphics[height=1.7in, angle=-90]{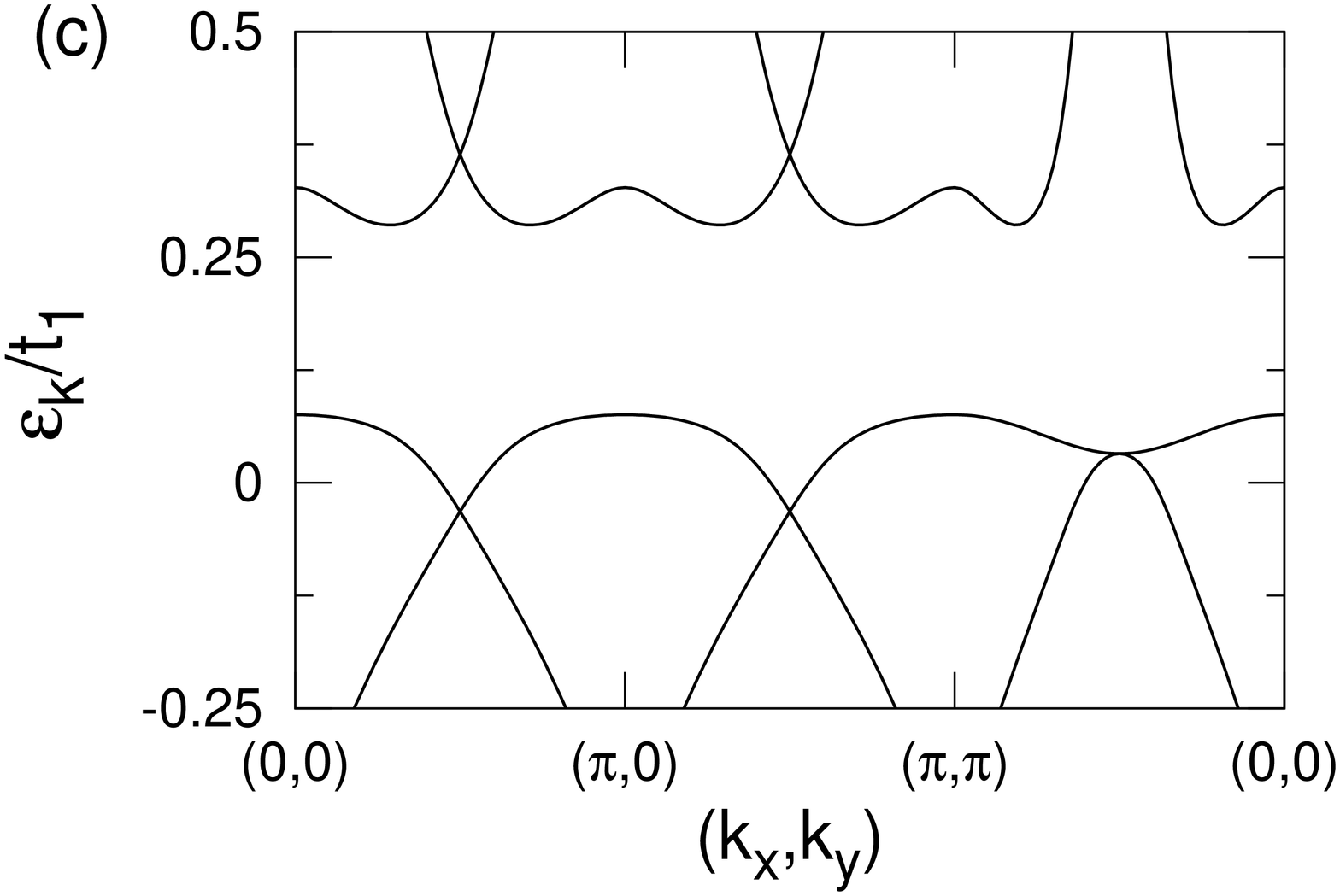}
\end{minipage}
\caption{(color online).
(a)
The bond and Kondo singlet parameters
at a fixed ratio $J_2/J_1=2.0$ as a function of $J_K$,
displaying a first order transition from the SSL-VBS to the HFL phase.
We  
%have rescaled the
%parameters so that they are dimensionless and 
show the three dimensionless independent quantities
for the solution that breaks no lattice translational symmetry.
Band structure along high symmetry directions in the reduced Brillouin zone in the SSL-VBS phase
with $J_K=0$ and $J_2/J_1=2.0$ (b) and in the HFL phase with $J_K=2.1t_1$ and $J_2/J_1=2.0$ (c).
The thick blue lines are the gapped spinon dispersion in the SSL-VBS phase.
}
\label{fig:b-structure}
\end{figure}

The zero temperature phase diagram
is shown in figure \ref{fig:phase-diagram}.
Without loss of generality, we have chosen $t_1/t_2=1.0$ and $n_c=0.5$
for
Figs.~\ref{fig:phase-diagram} and
\ref{fig:b-structure}.
For small Kondo coupling and a large $J_2/J_1$ ratio,
a VBS ground state
arises for which only $Q_{x+y}=Q_{x-y}$ are nonzero.
The singlet bonds are the same as in the pure Heisenberg model in the Shastry-Sutherland lattice
at large $J_2/J_1$, and we label it as
SSL-VBS.
This solution does not break any symmetry of the SSL.

Keeping $J_K/t_1$ small and decreasing $J_2/J_1$, we find a first order
transition at $J_2/J_1=1$ from the SSL-VBS to a plaquette VBS (P-VBS)
ground state where only
$Q_{x2}=Q_{x4}=Q_{y1}=Q_{y2}$ are nonzero. The P-VBS ground state
breaks a reflection
symmetry about either of the diagonal bonds in the the SSL.
It is degenerate with the conventional VBS on the square lattice
with only $Q_{x1}=Q_{x4}$ being nonzero.

%%%%%%%%%%%%%%%%%%%% Large-N phase diagram %%%%%%%%%%%%%%%%%%%%%%%%
%%%%%%%%%%%%%%%%%%%% Large-N phase diagram %%%%%%%%%%%%%%%%%%%%%%%%

For a large Kondo coupling we find a heavy Fermi liquid (HFL)
ground state,
which has a nonzero Kondo parameter $b_A=b_B=b_C=b_D $.
The singlet bond parameters are also nonzero: $Q_{xi}=Q_{yi}$, for $i=1-4$ and $Q_{x+y}=Q_{x-y}$.
We also obtain $\lambda_A=\lambda_B=\lambda_C=\lambda_D$, so the solution
does not break any symmetry of the SSL.
Here, we
 find that each $Q_{ij}$ acquires a finite phase $Q_{ij}=|Q_{ij}|e^{i\phi_{ij}}$. Correspondingly,
we define a gauge independent flux through
the triangular and square plaquettes as $\Phi_{\triangle}=\sum_{\triangle} \phi_{ij}(\mathrm{mod}\,\,2\pi)$ and
 $\Phi_{\square}=\sum_{\square} \phi_{ij}(\mathrm{mod}\,\,2\pi)$, respectively, where
 the summation is over the bonds around
 a plaquette.
 For the range of fillings $0<n_c \lesssim 0.75$, we find $\Phi_{\triangle}=\pi$ and $\Phi_{\square}=0$,
 whereas for $0.75\lesssim n_c<1$ we obtain $\Phi_{\triangle}=0$ and $\Phi_{\square}=0$.
 The finite flux through each triangular plaquette is a consequence of the spinons acquiring
 a finite kinetic energy
 from  their hybridization with the conduction-electron band;
 we can therefore consider this as
 a hybridization induced flux phase.
 However, even though the flux through each triangular plaquette is $\pi$,
 the total flux through each square plaquette is still zero (mod. $2\pi$);
 the flux
 does not affect the electronic band structure in the HFL phase.

We now turn to the transition among the two VBS phases
and the HFL phase.
Restricting the
solution to these three states,
we obtain the phase boundary in Fig.~\ref{fig:phase-diagram}
and the mean field parameters shown in Fig.~\ref{fig:b-structure}(a).
Unexpectedly,
 when considering the general solution
 we find
a number of intermediate states that
break the lattice symmetry,
in the region shown as the grey shaded area in Fig.~\ref{fig:phase-diagram}.
In some cases, for example the intermediate phase between the SSL-VBS phase
and the HFL phase
we find a 
PKS state:
some (half) of the moments in the unit cell are still locked into valence bonds,
while the other spins are
Kondo screened. This is discussed in detail
in the supplementary material~\cite{supp}.
Tuning $t_1/t_2$ only affects the location of the phase boundary;
a smaller ratio of $t_1/t_2$ makes 
%it easier for the Kondo singlets to form and therefore 
the transition
between each VBS phase and 
the HFL phase occur for smaller values of $J_K/t_1$.

%%%%%%%%%%%%%%%%%%%% mean field params and band structure %%%%%%%%%%%%%%%%%%%%%%%%
%%%%%%%%%%%%%%%%%%%% mean field params and band structure %%%%%%%%%%%%%%%%%%%%%%%%

\emph{Magnetism at $N=2$}:
We now incorporate long range AF
order into
our approach.
To do so we decouple the Heisenberg term into two distinct channels,
but we no longer have access to the large-$N$ limit and
are restricted
to $N=2$.  
In keeping with the generalized procedure of Hubbard-Stratonovich decouplings~\cite{NG.book},
and similar to reference~\cite{Senthil}, we rewrite the Heisenberg term in Eq.~(\ref{eqn:ham}) 
as follows: $J_{ij}{\bf S}_i\cdot{\bf S}_j
=xJ_{ij}{\bf S}_i\cdot{\bf S}_j+(1-x)J_{ij}{\bf S}_i\cdot{\bf S}_j$; the term proportional to $x$
is treated
within the RVB decoupling described previously.
The additional term
is decoupled in terms of N\'eel
order:
$(1-x)J_{ij}{\bf S}_i\cdot{\bf S}_j=(1-x)J_{ij}(2{\bf M}_i\cdot{\bf S}_j  -{\bf M}_i\cdot{\bf M}_j)$,
where $ {\bf M}_i=\langle {\bf S}_i \rangle$.
We consider the N\'eel
ground state with an ordering wave vector ${\bf Q}=(\pi/a,\pi/a)$ .
This AF order corresponds to ${\bf M}_A = {\bf M}_D = -{\bf M}_C =-{\bf M}_B={\bf M}$
within the four site unit cell.  
In the absence of a Kondo coupling, $J_K=0$, the phase diagram of the Heisenberg model
as a function of $x$ and $J_2/J_1$ is presented in the supplementary material~\cite{supp}.

The phase diagram at $J_K=0$ provides the physical basis for choosing the parameter $x$.
Within our fermonic representation,
classical AF order arises at $x=0$, corresponding to a full ordered magnetic moment $|{\bf M}|=1/2$.
%with gapped non-dispersive excitations.  
The ground state wave function of the true quantum AF state is known to contain considerable RVB correlations~\cite{Dagatto},
suggesting a choice of  $x>1/2$.
Indeed, for the nearest neighbor Heisenberg model ($J_K=0$ and $J_2=0$) we find a quantum
AF phase as a self consistent solution for $x$ in the range $0.67 \leq x < 0.8125$, which has a lower free energy
than the classical N\'eel state
($Q_{xi}=Q_{yi}=0$).
This state, is a free energy local minimum,
and taken as a candidate of the true N\'eel ground state.
The quantum AF phase
has finite RVB 
%singlet 
parameters  
$Q_{xi}=Q_{yi}$ for $i=1-4$,
which reduce 
the ordered moment $|{\bf M}|<1/2$.
%, while making the excitations dispersive.
Incorporating fluctuations further will reduce the free energy even more, making
the AF phase the true ground state in the limit $J_2/J_1\ll1$.
Here we present the results for $x=0.7$.  The phases and the overall profile of the phase diagram are not sensitive to
the choice of $x$ in the range $0.67 \leq x < 0.8125$, as discussed in the supplementary material~\cite{supp}.

The resulting phase diagram is given in Fig.~\ref{fig:M-pd}, for parameters $n_c=0.5$ and $t_1/t_2=1$.
We have restricted
the solutions
to states that do not break any lattice symmetries.
For small $J_K/t_1$ and tuning the ratio of $J_2/J_1$,
we find a first order transition from the AF phase to the SSL-VBS phase.
For small $J_2/J_1$, and tuning the Kondo coupling, the AF phase has a
continuous transition~\cite{note} into a spin density wave (SDW) phase
characterized by the onset of Kondo screening:
$b_A=b_B=b_C=b_D$ increases
continuously from zero with
nonzero values of ${\bf M}$, $Q_{xi}=Q_{yi}$ for $i=1-4$ and
$Q_{x+y}=Q_{x-y}$.  
%\JP{We stress that we have found the SDW phase can only be stabilized with \emph{finite} RVB parameters. Therefore, incorporating the %quantum fluctuations about the classical N\'eel ground state and the subsequent reduction of the ordered moment is essential for the local %moments to participate in both magnetic order and singlet formation with the conduction electrons.}  
Upon increasing $J_K$ further,
there is a first order transition from the SDW phase into the HFL phase with ${\bf M}=0$.

Our results demonstrate a rich interplay between
Kondo and RKKY interactions.
In addition to the AF order and its suppression, there is also the competition between
the Kondo effect and VBS order in the magnetically-disordered region.
In the notation of Fig.~\ref{fig:model}(a), we associate Kondo hybridization ($b\ne 0$)
with a large Fermi surface (subscript
$L$) and Kondo destruction ($b =0$) with a small Fermi surface (subscript $S$).
The phase diagram we have calculated, Fig.~\ref{fig:M-pd}, represents a remarkable realization
of the global phase diagram that had been advanced on qualitative considerations \cite{Si2,Coleman}.
It will be instructive to study Kondo lattice models in other geometrically frustrated cases,
for example Kagome lattices (pertinent to CePd$_{1-x}$Ni$_x$Al~\cite{Fritsch})
and triangular lattices (relevant to YbAgGe~\cite{Morosan} and YbAl$_3$C$_3$~\cite{Khalyavin}),
and explore the generality of the global phase diagram.
Compared to those cases, the Kondo model on the SSL has
the main advantage that the magnetically frustrated regime is accessible by a large-$N$ approach.

Several remarks are in order. First,
 in the phase diagram of Fig.~\ref{fig:M-pd}, we find a \emph{line}
 of direct
 transitions from AF$_S$ to P$_L$.
However,
whether this is a line of transitions or a single point is sensitive to the model parameters
in our approach
and for $x=0.75$ we find the transition collapsing to a single point (see the supplementary material~\cite{supp}).
It is
important
to consider how further quantum fluctuations
will affect the topology of the phase diagram in Fig.~\ref{fig:M-pd}.
Recently, insights have been gained from calculations on a quantum impurity model incorporating
local quantum fluctuations~\cite{Nica}; within an extended dynamical mean field context~\cite{LCP},
the results of Ref.~\cite{Nica} imply this direct transition to be
a line in the phase diagram.

Second, due to an even number of spins per unit cell,
the spinon bands are either empty or completely
full~\cite{Coleman}. Hence the
volume of the Fermi surface will not change when the system goes from the
SSL-VBS to the HFL phase. Nonetheless, the topology of the
Fermi surfaces reflects the incorporation ($L$) or absence ($S$) of the Kondo resonances
in the Fermi volume and can be different for the two cases.  We show the Fermi surfaces
for $n_c=0.5$ in the supplementary material~\cite{supp}.

Third, it is instructive to compare our results to those of Ref.~\onlinecite{Bernhard}. Where there is overlap,
the results of that work and ours are largely consistent. We are able to draw substantially new implications by studying
the competitions of all the phases pertinent to the global phase diagram, including the $AF_L$ phase.
Furthermore, our work has also uncovered PKS phases.
In a similar vein, we note that the Shastry-Sutherland Kondo lattice was also considered in Ref.~\onlinecite{Coleman},
with a particular focus on possible superconducting pairings. The implications of our study
for superconductivity is an intriguing issue, but is beyond the scope of the present work.

%%%%%%%%%%%%%%% Model fig %%%%%%%%%%%%%
\begin{figure}[t]
\includegraphics[height=2.in]{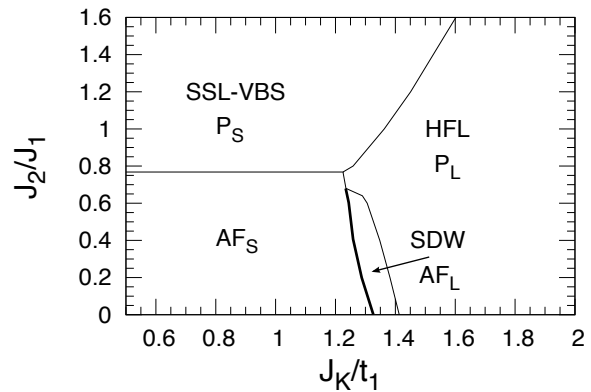}
\caption{Phase diagram of the
Shastry-Sutherland Kondo lattice incorporating magnetic order
for a metallic filling $n_c=0.5$.
Thin (thick) lines represent first order (continuous~\cite{note})  transitions.}
\label{fig:M-pd}
\end{figure}
%%%%%%%%%%%%%%% Model fig %%%%%%%%%%%%%

The systematic nature of our results is important not only for generating insights into the global phase diagram,
but also
to drawing implications 
for experiments in heavy-fermion metals.
Our phase diagram
in Fig.~\ref{fig:M-pd} opens up a trajectory from
the AF$_S$ to the HFL phase via a sequence of quantum phase transitions
that passes
through
a VBS, $P_S$ phase.
This result has implications for Yb$_2$Pt$_2$Pb, 
%which
%is metallic, and contains a sizable Kondo coupling as evidenced by the observation of a large
%specific-heat~\cite{Kim}.
where
experiments~\cite{Kim} appear
to have realized such a sequence of transitions [see figure~\ref{fig:model}(a)].
%; evidence for the
%VBS phase has
%come from the determination of the spin-dimerization gap.

In addition, we have provided evidence for
 intermediate, partially Kondo screened phases
 {\it in metallic cases.}
This type of phase has also been discussed in a variational quantum Monte Carlo approach
in Kondo insulator settings
~\cite{Motome}.
In this regard, it is intriguing that
experiments on the metallic
CePd$_{1-x}$Ni$_x$Al~\cite{Fritsch} have suggested that the frustration
in this material
is not large enough to yield a spin liquid,
but instead
leads to a ground state
where some of the
magnetic moments
form long range
AF order, while the others are completely screened by the Kondo effect.

In conclusion, we have studied the global phase diagram in the prototypical geometrically-frustrated
Shastry-Sutherland Kondo lattice.
Our work represents the first concrete calculation in which all four phases, AF$_S$, AF$_L$, P$_S$, and P$_L$
%with and without static
%Kondo screening
%and in the presence and absence of antiferromagnetic order, 
appear in a single zero-temperature
phase diagram.
Our results have elucidated the rich variety of quantum phases and their transitions in heavy-fermion metals,
and provide new insights into the puzzling experimental observations recently made
in geometrically frustrated heavy fermion metals.
%such as Yb$_2$Pt$_2$Pb and CePd$_{1-x}$Ni$_x$Al.

\emph{Acknowledgements}: We would like to
acknowledge useful discussions with
D. T. Adroja,
M. Aronson, C. H. Chung, P. Coleman, S. Kirchner, A. Nevidomskyy,  and
E. Nica.
This work was supported in part by
the NSF Grant No.\ DMR-1309531, the Robert A.\ Welch Foundation Grant No.\ C-1411,
the Alexander von Humboldt Foundation and the East-DeMarco fellowship (JHP).
The majority of the calculations have been performed on the Shared University Grid at Rice funded by NSF
under Grant EIA-0216467, and a partnership between Rice University, Sun Microsystems,
and Sigma Solutions, Inc..
Q.S.\  also acknowledges the hospitality of the the Karlsruhe Institute of Technology,
the Aspen Center for Physics (NSF Grant No.\ 1066293),
and the Institute of Physics of Chinese Academy of Sciences.

%%%%%%%%%%% Supplementary Material %%%%%%%%%%%%%
\newpage
\onecolumngrid
\setcounter{figure}{0}
\makeatletter
\renewcommand{\thefigure}{S\@arabic\c@figure}
\setcounter{equation}{0} \makeatletter
\renewcommand \theequation{S\@arabic\c@equation}

\section*{{\Large Supplementary Material } \\
Quantum phases of the Shastry-Sutherland Kondo lattice:\\
implications for the global phase diagram of heavy fermion metals}
{\hfill J.\ H.\ Pixley, Rong Yu and Qimiao Si \hfill}
\vskip 1.0 cm
\section{Numerical Method to Solve the Self-Consistent Equations}
In this work, we minimize the ground-state energy with respect to the self-consistent parameters $Q_{ij}$, $b_X$, and $\lambda_X$ by solving the coupled non-linear equations. To solve the equations efficiently, we apply the Broyden's mixing method~\cite{Broyden65,NumRecp07,Johnson88}, which has been widely used in density functional theory (DFT)~\cite{MarksLuke08,Baran}, in dynamical mean-field theory~\cite{Zitko09}, and in other contexts of correlated electron problems~\cite{Yunoki08,Yu13}. 
In general, the method solves the self-consistent equations $\boldsymbol{X}=\boldsymbol{F}[\boldsymbol{X}]$ iteratively with a set of initial values of the parameters $\boldsymbol{X}^{(0)}_{\mathrm{in}}$. At the $m$-th iteration, a new set of the parameters is obtained via
\begin{equation}
 \boldsymbol{X}^{(m)}_{\mathrm{out}} = \boldsymbol{F}[\boldsymbol{X}^{(m)}_{\mathrm{in}}].
\end{equation}
The input of the next iteration is constructed from 
\begin{equation}
 \boldsymbol{X}^{(m+1)}_{\mathrm{in}} = \boldsymbol{X}^{(m)}_{\mathrm{in}} - \boldsymbol{B}^{(m)} (\boldsymbol{X}^{(m)}_{\mathrm{out}} - \boldsymbol{X}^{(m)}_{\mathrm{in}}),
\end{equation} 
where $\boldsymbol{B}^{(m)}$ is an approximation to the inverse Jacobian matrix of the non-linear equations at the $m$-th iteration. The Broyden's method provides a scheme to update the approximate inverse Jacobian $\boldsymbol{B}^{(m)}$ at each iteration such that good convergence is obtained~\cite{Broyden65,NumRecp07,Johnson88}.

To achieve the global minimum of the ground-state energy, we have done a random search of initial conditions in the parameter space. The typical number of the $\boldsymbol{X}^{(0)}_{\mathrm{in}}$ configurations used in the calculation is $10^3$. For each given initial condition, 
a Broyden mixing is performed; when convergence is approached, the corresponding ground-state energy is recorded. 
The global minimum of the ground-state energy is then obtained by comparing the energies for different configurations.

\section{Intermediate Phases and the Effect of Model Parameters}
\begin{figure}[!h]
\includegraphics[height=2.0in]{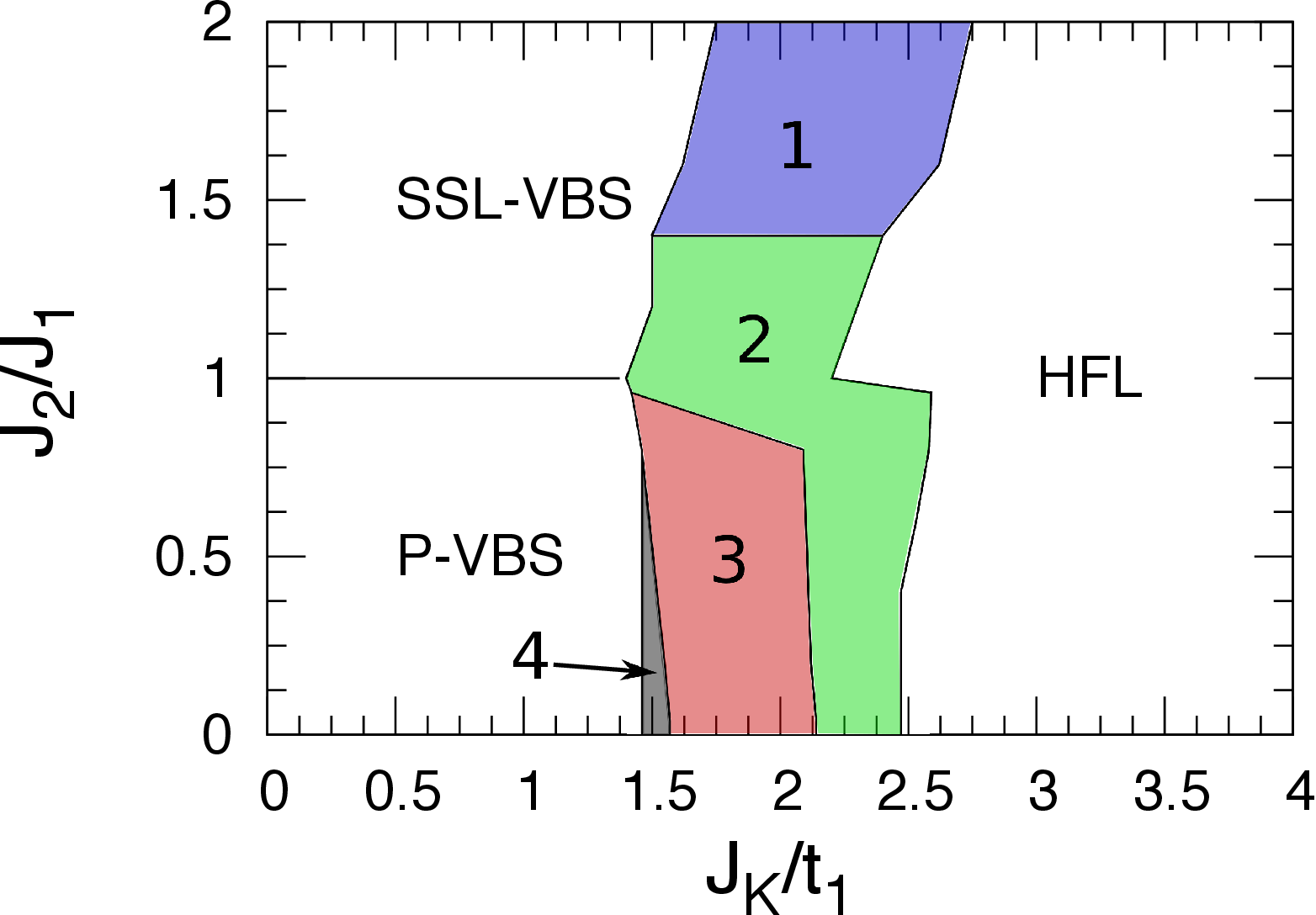}
\caption{Full large-N phase diagram of the SS Kondo lattice, for $t_2=t_1$ and $n_c=0.5$. The phases are 
described in the text. }
\label{fig:N-pd}
\end{figure}

Here we present the full Large-N phase diagram for the metallic SS Kondo lattice, focusing in particular
on the intermediate phases sketched in Fig.~2 of the main text.  We consider tight binding parameters 
$t_1=t_2=1$, and a conduction electron filling $n_c=0.5$.  We find a series of intermediate states that
break the 
lattice symmetry within the unit cell.  In particular we find four intermediate states labelled with different colors
in Fig.~\ref{fig:N-pd}: blue (1), green (2), red (3) and grey (4). 

We first consider the blue region (1), which is the most physically relevant intermediate phase because it 
occurs in the transition region between the SSL-VBS and HFL phases.  
Here, we find a phase with \emph{partial Kondo screening}
defined as $b_A=b_D> 0$, $b_B=b_C=0$, with $Q_{x-y} \gg Q_{x+y}  > 0$. The local moments
on sublattices A and D are screened by the Kondo effect, whereas those on the $B$ and $C$
sub-lattices are locked into a VBS singlet.
 Turning next to the green region (2), we find a square plaquette RVB phase with Kondo screening, 
 defined by $b_A=b_D>b_B=b_C>0$, $Q_{x-y}>Q_{x+y}>0$, $Q_{x1}=Q_{y1}=Q_{x3}=Q_{y2}$ 
 and $Q_{x2}=Q_{y3}=Q_{x4}=Q_{y4}$.  In this phase, plaquette valence bonds form on each square plaquette
 that contains a $J_2$ bond.  
The red region (3) describes a phase where the RVB parameters are in a ``kite'' phase, and the local moments
are screened. It is defined by $b_A=b_D>b_B=b_C$, $Q_{x+y}\neq Q_{x-y}$, $Q_{x1}=Q_{y1}$, 
$Q_{x2}=Q_{y3}$, $Q_{x3}=Q_{y2}$, and $Q_{x4}=Q_{y4}$. 
Lastly, we come to the  grey region (4), which corresponds to a spin-Peierls  phase with partial Kondo screening.
Here,  $b_A=b_B \neq 0$, $b_C=b_D=0$ with $Q_{x3}\neq0$; all the other parameters vanish.

%%%%%%%%%%%%%%% Model fig %%%%%%%%%%%%%
%%%%%%%%%%%%%%% Model fig %%%%%%%%%%%%%

%%%%%%%%%%%%%%% Model fig %%%%%%%%%%%%%
\begin{figure}[!h]
\begin{minipage}[b]{20pc}
\includegraphics[height=2.0in]{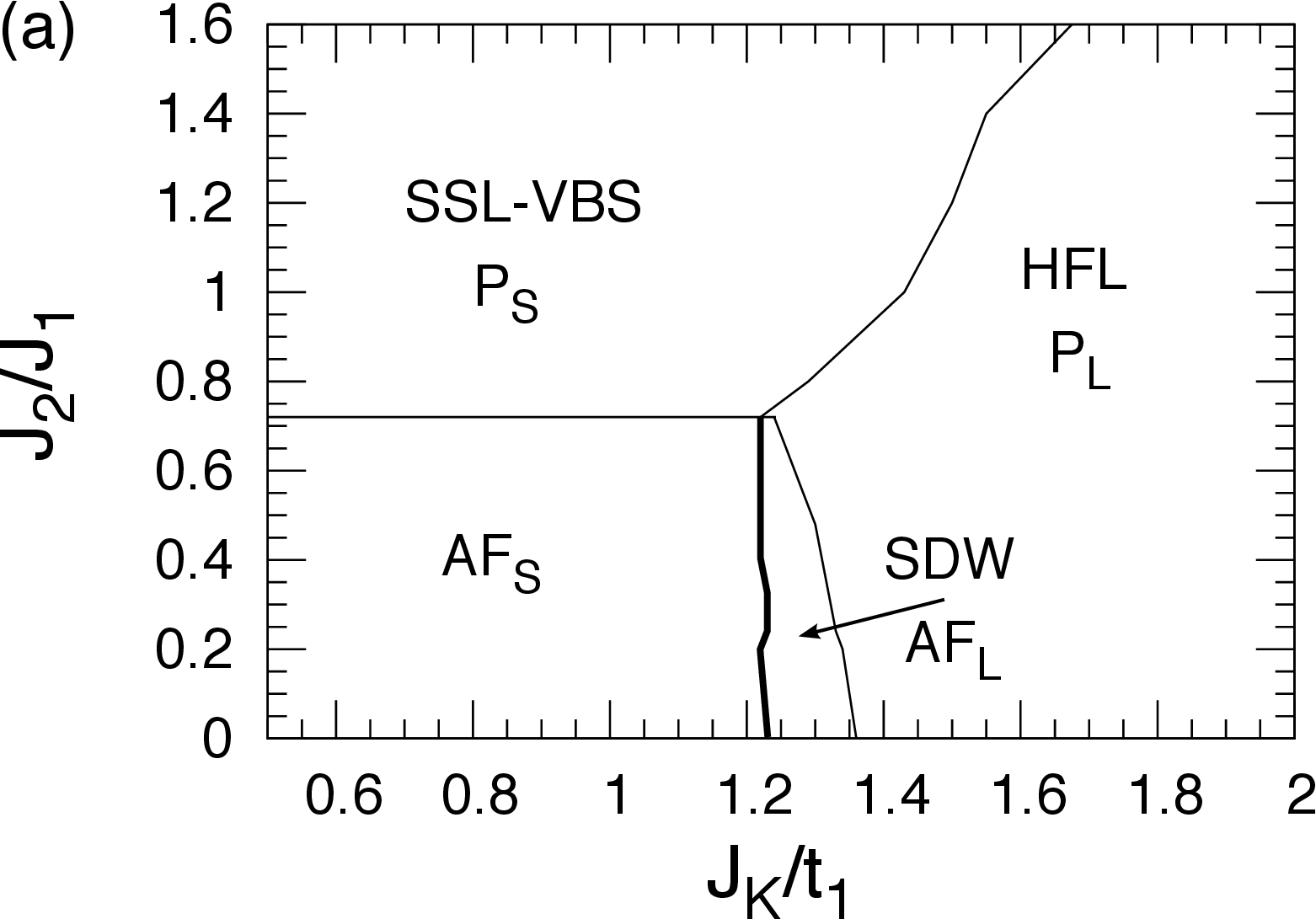}
\end{minipage}
\begin{minipage}[b]{20pc}
\includegraphics[height=2.0in]{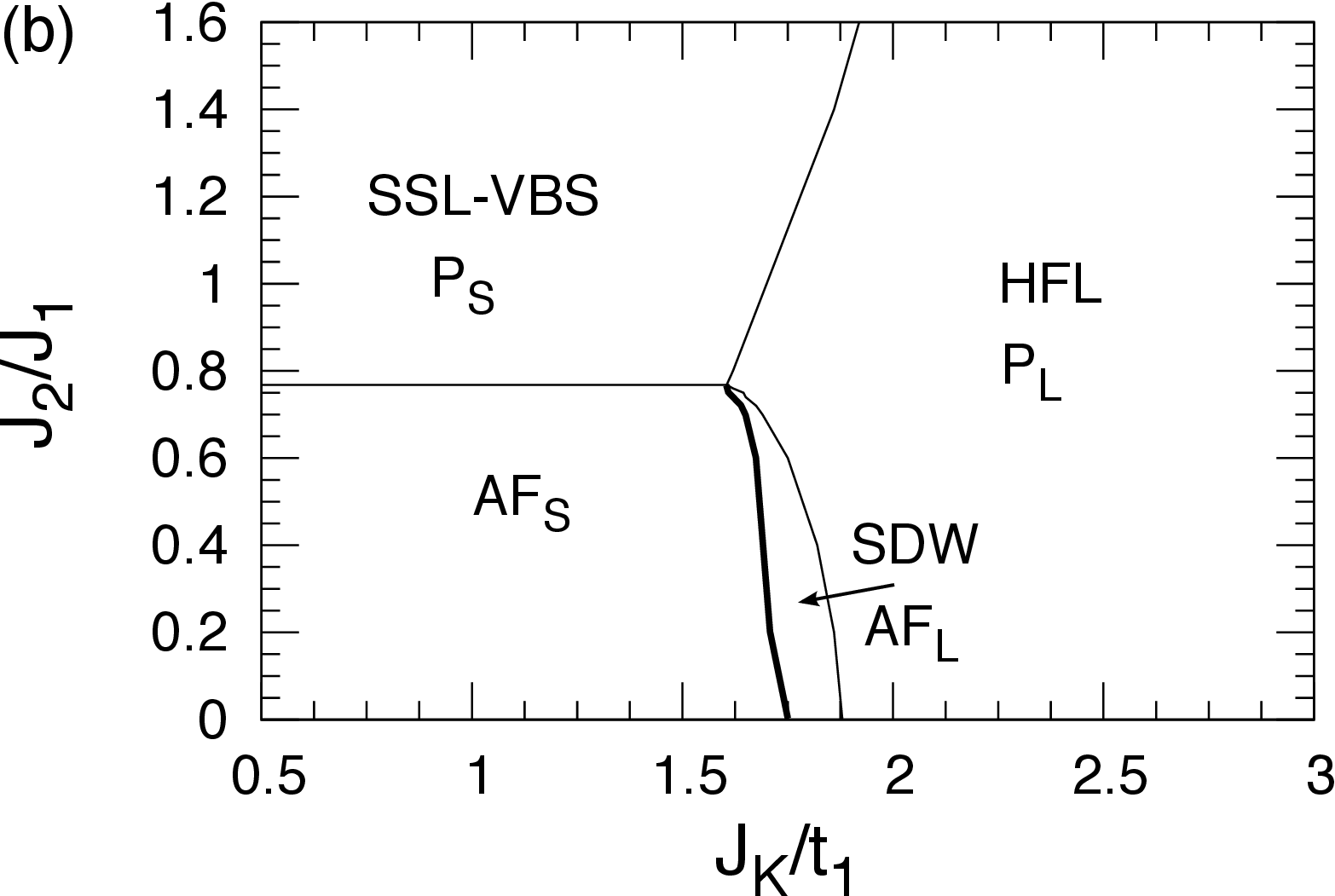}
\end{minipage}
\caption{Phase diagram of the SS Kondo lattice incorporating magnetic order using $x=0.75$, while keeping
$t_2=t_1$ (a) and for $x=0.70$ with $t_2=0$ (b).  In both cases, the conduction electron filling, $n_c=0.5$,
is unchanged from before.
The thin (thick) lines represent first order  (continuous) transitions.}
\label{fig:N-pd2}
\end{figure}
%%%%%%%%%%%%%%% Model fig %%%%%%%%%%%%%

We now turn the effect of changing model parameters on the global phase diagram.
We first consider choosing a different value of $x=0.75$ (see the main text for the definition of $x$) while
keeping $t_1/t_2=1$ and $n_c=0.5$.  
This choice of $x$ is still in the range over which a quantum antiferromagnetic phase arises (see below, Fig.~\ref{fig:M-pd}).
Separately, 
we consider keeping $x=0.70$ but setting $t_2=0$, with $t_1/J_1=4.0$ and $n_c=0.5$;
this allows us to study the effect of the conduction-electron band dispersion 
({\it cf.} Figs.~\ref{fig:t2=0_JK=0} and \ref{fig:t2=0_JK=2}).
Interestingly, for both cases we find the line of transitions between AF$_S$ and HFL collapse to a point.
Therefore, whether the transition between AF$_S$ and HFL is a line or a single point is a question
that needs to be addressed beyond the mean field level, for example within an extended dynamical
mean field approach. The key point, however, is that
the overall profile of the global phase diagram is robust against these
changes of parameters.

%%%%%%%%%%%%%%% Fermi Surface %%%%%%%%%%%%%
\begin{figure}[!h]
\begin{minipage}[b]{20pc}
\includegraphics[height=2.0in]{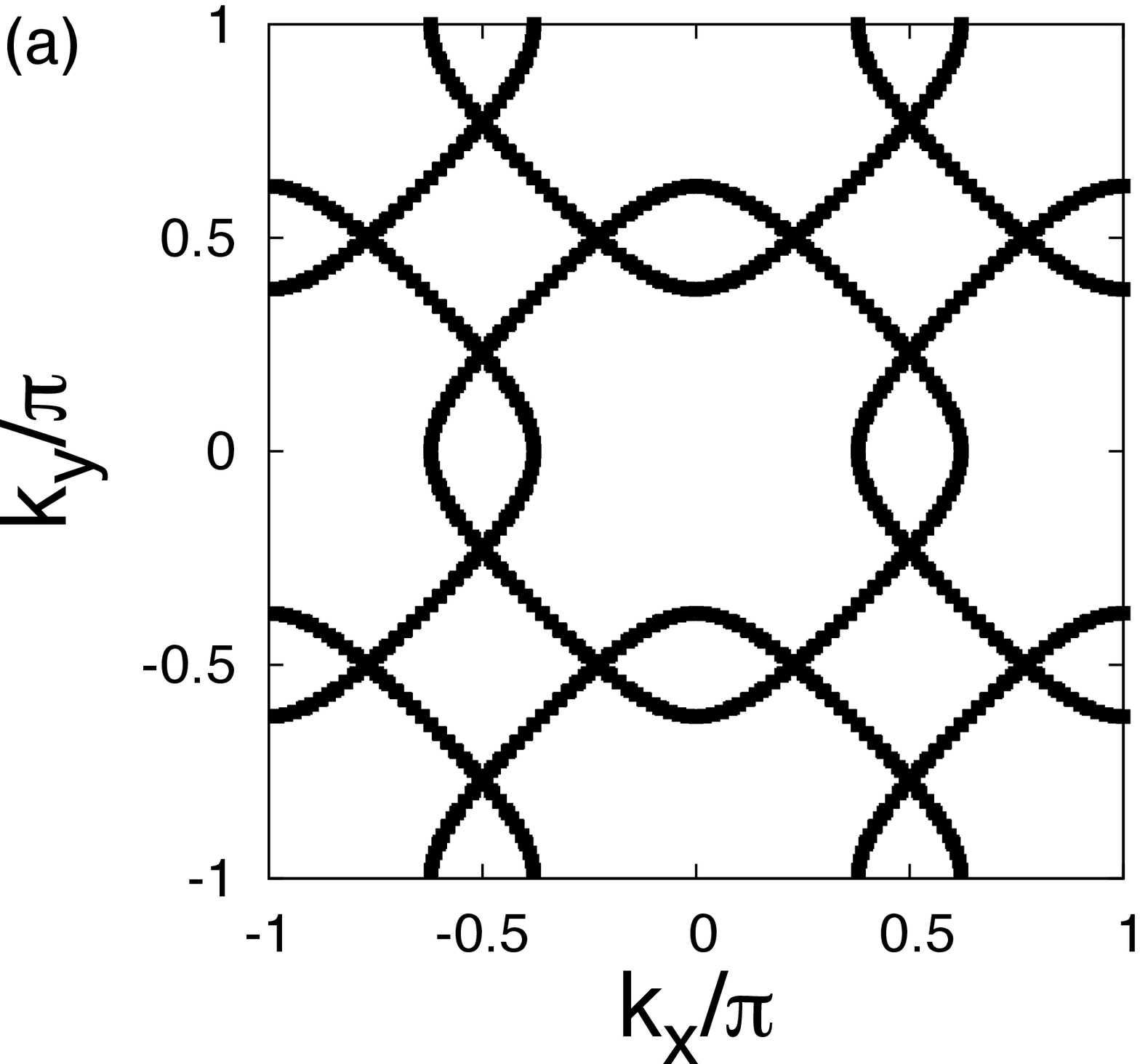}
\end{minipage}
\begin{minipage}[b]{20pc}
\includegraphics[height=2.0in]{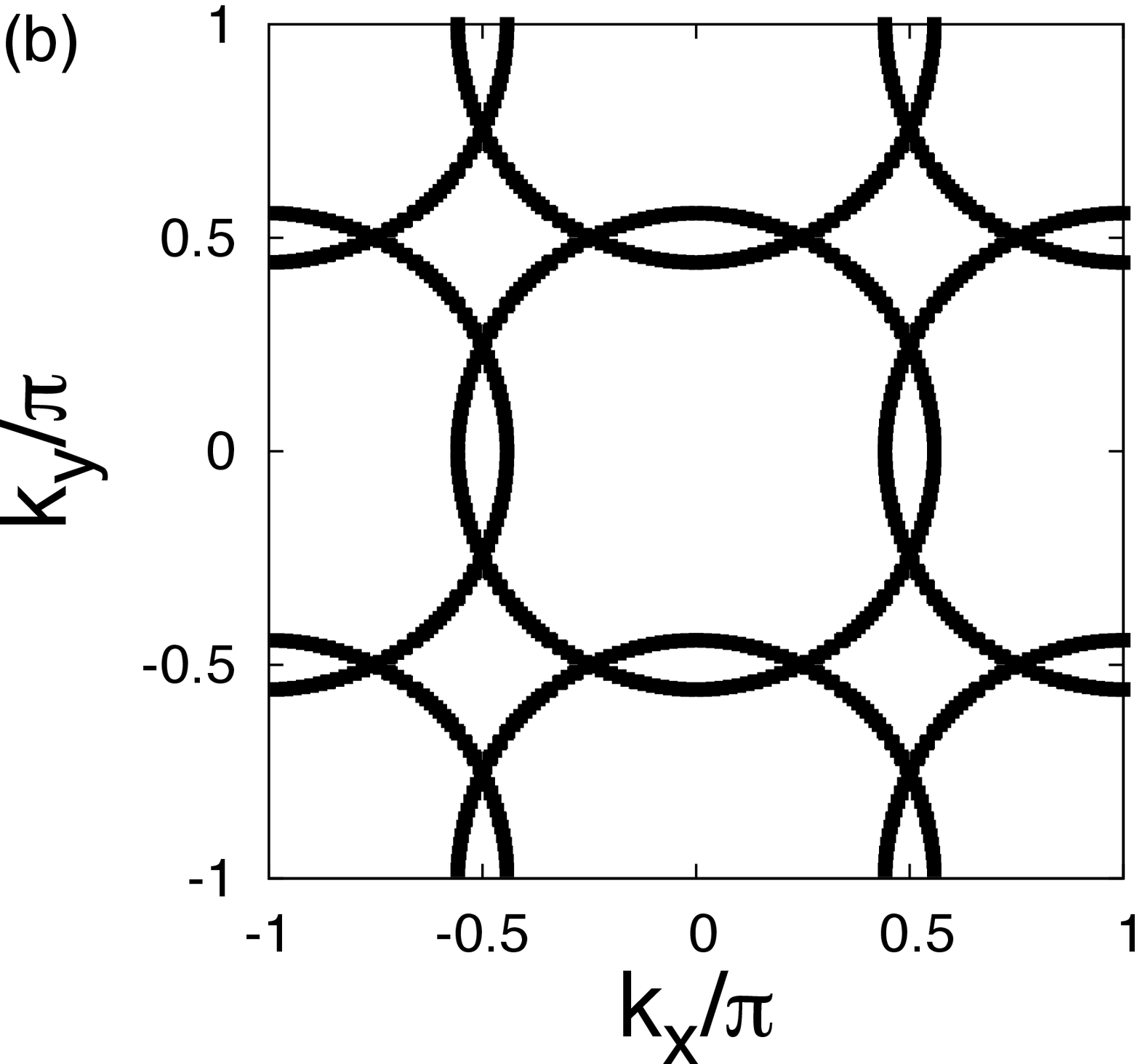}
\end{minipage}
\caption{(a) Fermi Surface corresponding to the band structure in Fig.~3(b) of the main text, in the VBS phase.
Here, $J_2/J_1=2$, $J_K/t_1=2.1$. Since the spinon bands are gapped, this corresponds to the band structure
of the conduction electron dispersion alone, defined on the SSL with $t_1/t_2=1$ and $n_c=0.5$; 
(b)  Fermi Surface corresponding to the band structure
in Fig.~3(c) of the main text, in the HFL phase. The bare parameters are the same as in (a).
Note that the Fermi volume does not change from (a) to (b), due to the even number of spins per unit cell 
as described in the main text.}
\label{fig:N-pd3}
\end{figure}
%%%%%%%%%%%%%%% Model fig %%%%%%%%%%%%%

To explore further the robustness of our results against the change of the conduction-electron dispersion,
we also discuss the effect of an additional tight binding parameter $t_3$, which connects every next nearest
neighbor that is not connected via $t_2$.  Just like tuning $t_2/t_1$ away from $1$, any
 finite $t_3$ will introduce a curvature to the region of the flat band; this flat portion
existed for $t_2/t_1=1$ (and $t_3=0$), along
the $k_x=k_y$ direction in the Brillouin zone and away from the Fermi energy, as shown in Fig.~3(b) in the main text. 
We find that tuning the ratio
of $t_3/t_1$ can change the degree to which the intermediate PKS phases occur.
The region of the intermediate phases that break the lattice symmetry within the unit cell narrows
for increasing $t_3/t_1$, and can even be completely eliminated for a large $t_3/t_1$ ratio. 
Again, the overall profile of the global phase diagram is robust against the change of $t_3/t_1$. 

We close this subsection by showing the Fermi surfaces of the SSL-VBS and HFL phases, both for the 
case $t_2=t_1$ and $n_c=0.5$ (Figs.~\ref{fig:N-pd3}(a) and (b), respectively) and for
$t_2=0$ and $n_c=0.5$ (Figs.~\ref{fig:t2=0_JK=0} and \ref{fig:t2=0_JK=2}, respectively).

%%%%%%%%%%%%%%% Fermi Surface %%%%%%%%%%%%%
\begin{figure}[!h]
\begin{minipage}[b]{20pc}
\includegraphics[height=2.0in]{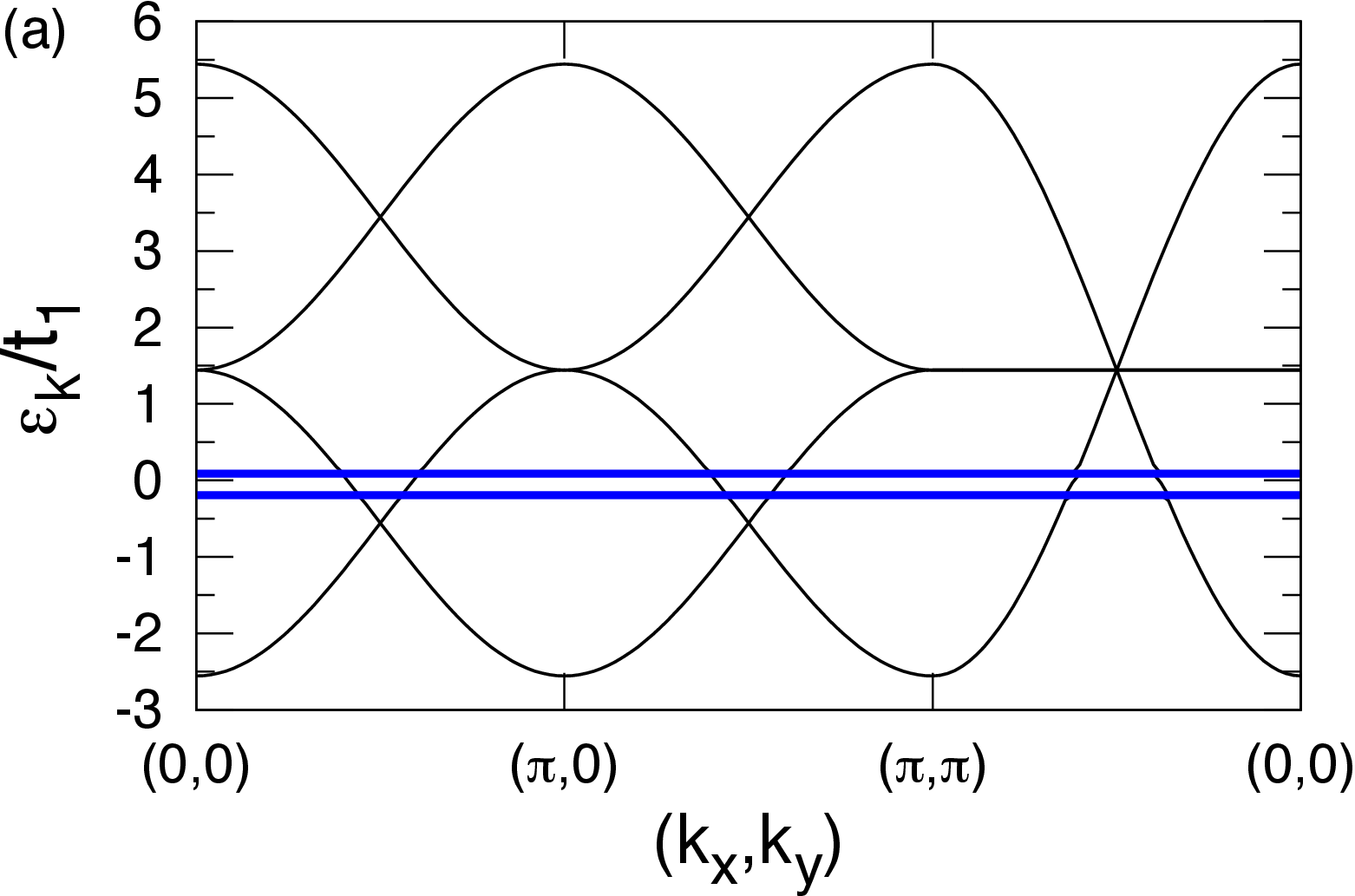}
\end{minipage}
\begin{minipage}[b]{20pc}
\includegraphics[height=2.0in]{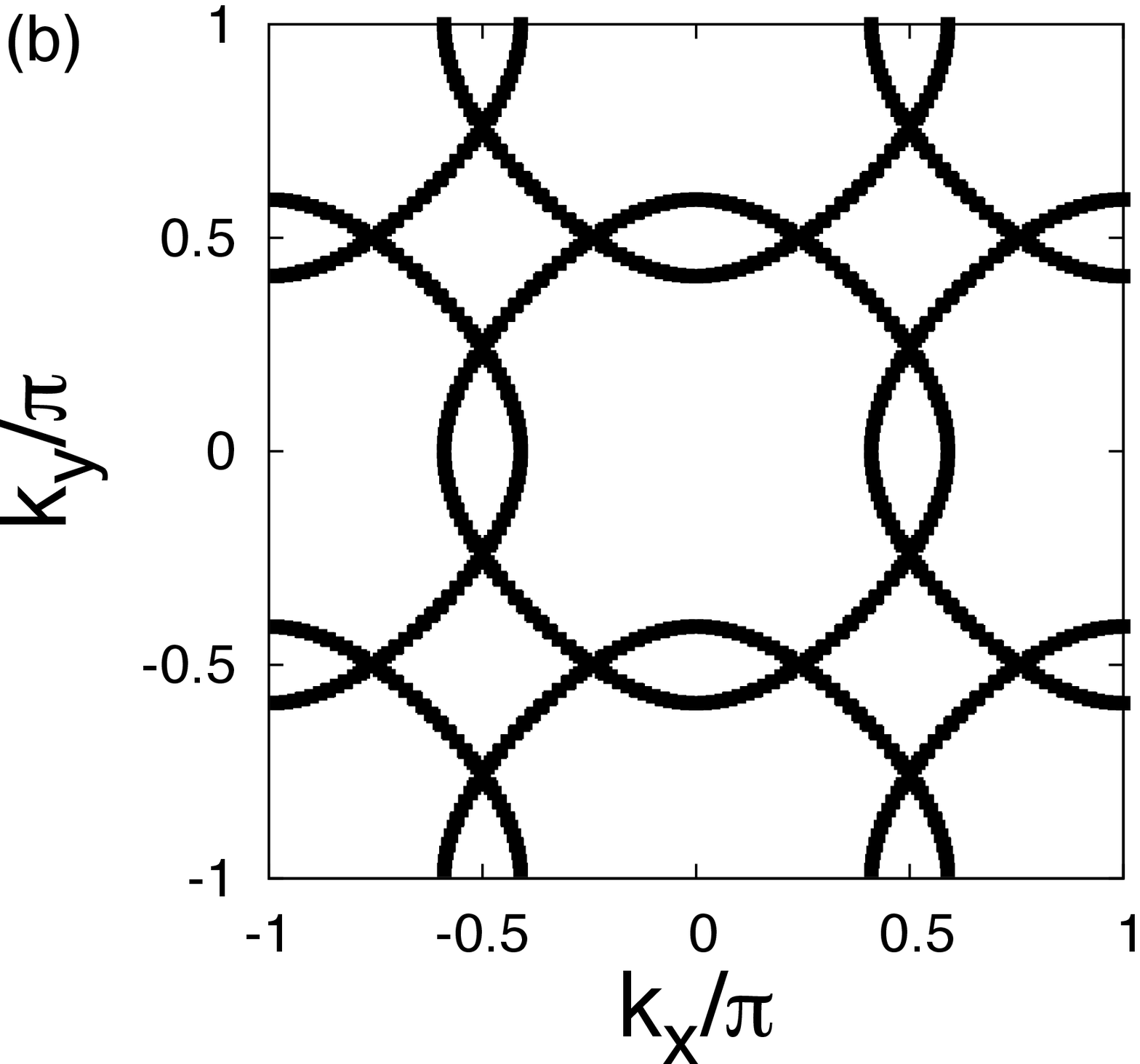}
\end{minipage}
\caption{(a) Band structure in the VBS phase when the diagonal hopping vanishes, $t_2/t_1=0$,
and for $J_2/J_1=1.6$,  $J_K/t_1=0$ and $n_c=0.5$. The parameter $t_1=1.0$ sets the unit of energy.
The blue lines are the gapped spinon bands.  
The Fermi energy is at $\epsilon_k/t_1=0$;
(b) The corresponding Fermi surface.  }
\label{fig:t2=0_JK=0}
\end{figure}
%%%%%%%%%%%%%%% Model fig %%%%%%%%%%%%%

%%%%%%%%%%%%%%% Fermi Surface %%%%%%%%%%%%%
\begin{figure}[!h]
\begin{minipage}[b]{20pc}
\includegraphics[height=2.0in]{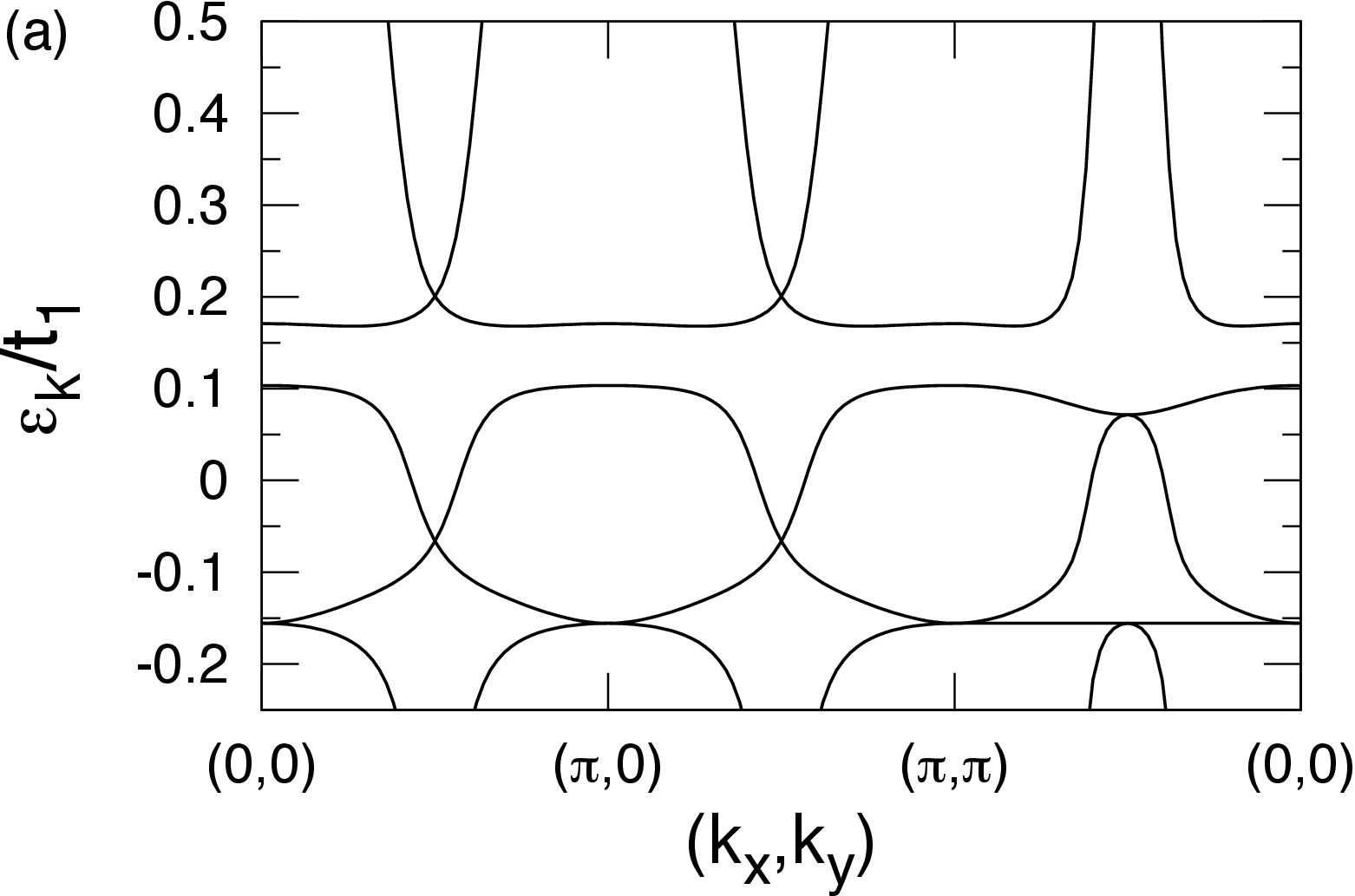}
\end{minipage}
\begin{minipage}[b]{20pc}
\includegraphics[height=2.0in]{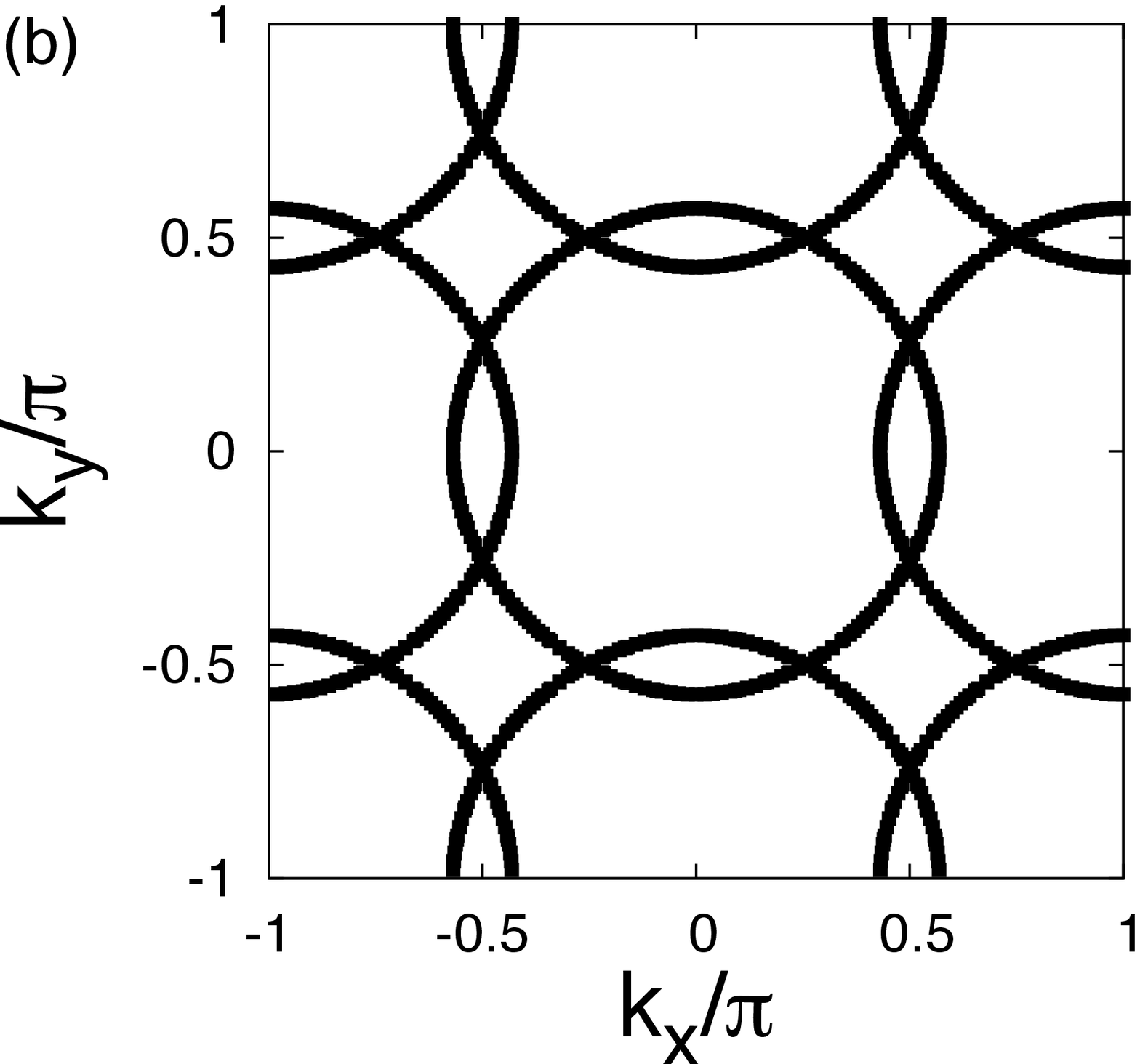}
\end{minipage}
\caption{(a) Band structure in the HFL phase, for parameters as in Fig.~\ref{fig:t2=0_JK=0}.
The fermi energy is also at $\epsilon_k/t_1=0$;
(b) The corresponding Fermi surface. }
\label{fig:t2=0_JK=2}
\end{figure}
%%%%%%%%%%%%%%% Model fig %%%%%%%%%%%%%

\section{Magnetic Phase Diagram}
Finally, we discuss the Heisenberg model in the absence of any Kondo coupling in our approach,
when both the RVB correlations and N\'eel order are incorporated and
the self consistent solutions with the lowest free energy are determined.
The region for the candidate quantum N\'eel phase, as described in the main text,
 is also considered as a self consistent solution. 
The resulting phase diagram is shown in Fig.~\ref{fig:M-pd}, where both the P-VBS and
 SSL-VBS phases have been described
in the main text.  The classical N\'eel state (Classical AFM) is defined as ${\bf M}_A = {\bf M}_D= -{\bf M}_B 
= -{\bf M}_C>0$, with all the other parameters equal to zero.  Here the order parameter retains the full classical moment.
We find the candidate quantum N\'eel phase, with ${\bf M}_A = {\bf M}_D= -{\bf M}_B 
= -{\bf M}_C>0$ and $Q_{xi}=Q_{yi}\neq0$, $
 i=1-4$  to be a self consistent solution only in the range $0.67 \leq x < 0.8125$.
 The nonzero RVB singlet parameters cause a reduced value for the ordered moment.
 This is shown as the magenta region in Fig.~\ref{fig:M-pd},
 where the boundary of the region at a finite ratio of $J_2/J_1$ marks the transition to the SSL-VBS phase.  
 For a judicious choice of $x$ in the range $0.6<x<0.67$ we find
 the transition from the classical AFM phase to the SSL-VBS has an intermediate P-VBS phase similar to what is found 
 in exact diagonalization studies (see reference [20] in the main text).  We remark that for $x=0.5$, 
 which treats N\'eel and VBS order on equal footing, we find the location of the transition from the classical AFM phase 
 to the SSL-VBS to be $(J_2/J_1)_c =1.35$ which is close to the value obtained from a variety of other methods 
 (see reference [21] in the main text).

%%%%%%%%%%%%%%% Mag PD fig %%%%%%%%%%%%%
\begin{figure}[!h]
\includegraphics[height=2.0in]{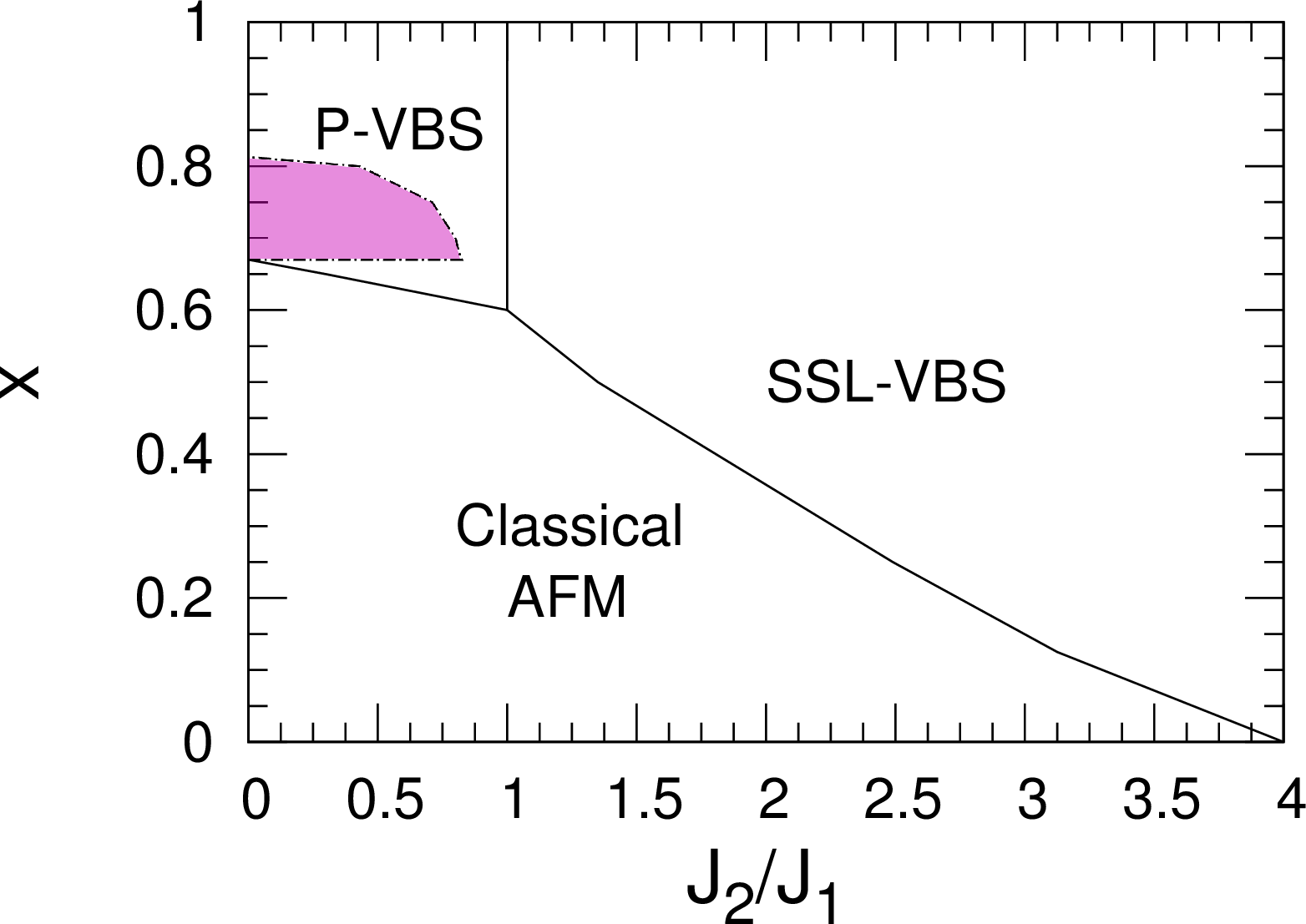}
\caption{The phase diagram of the Heisenberg model in the slave-fermion approach.
The parameter $x$ and the phases are described in the main text.}
\label{fig:M-pd}
\end{figure}
%%%%%%%%%%%%%%% Mag PD fig %%%%%%%%%%%%%


\begin{thebibliography}{1}
\bibitem{note}  We are not able to discern between a continuous and a very weakly first order transition.
\end{thebibliography}
\end{document}